\documentclass[12pt]{article}
\newcommand{\be}{\begin{equation}}
\newcommand{\bea}{\begin{eqnarray} \nonumber}
\newcommand{\ee}{\end{equation}}
\newcommand{\eea}{\end{eqnarray}}

 \newcommand{\EPJB}[3]{Eur.~Phys.~J.~B {\bf #1}, #2 (#3)}

\def\(({\left(}
 \def\)){\right)}
\def\[[{\left[}
\def\]]{\right]}
\def\bi{\bibitem}
\def \form#1 {eq. (\ref{#1}) }
\def \parziale#1#2  {{\partial {#1} \over \partial {#2}}}

\def \Tr {\mbox{Tr}}
\def \ba#1 {\overline{#1}}

\def  \si {\sigma}

\begin{document}

\title{ Two spaces looking for a geometer}

\author{Giorgio Parisi\\
Dipartimento di Fisica,  SMC and Udrm1 of INFM, INFN,\\
Universit\`a di Roma ``La Sapienza'',\\
Piazzale Aldo Moro 2,
I-00185 Rome (Italy)}
\maketitle

\begin{abstract}
In this talk I will introduces two {\sl spaces}: the first space is the usual $n$-dimensional vector 
space with the unusual feature that $n$ is non-integer, the second space is composed by the linear 
matrices acting on the previous space (physicists are particularly interested to study the limit 
$n$ going to zero).  These two {\sl spaces} are not known to most of the mathematicians, but they are 
widely used by physicists.  It is possible that, by extending the notion of space, they can become 
well defined mathematical objects.

\end{abstract}

\section{Introduction}

In last thirty years physicists have commonly  used {\sl spaces} having 
non-integer dimensions.  These {\sl spaces} do not have a clear mathematical meaning and 
it not evident in which sense the name space is not abusive.  On the other end it would be 
very interesting if one could properly define these objects, that are commonly used by 
physicists. It is possible this task may be achieved by generalizing the notion of space.

There are many examples of the previous construction. Here we will concentrate our 
attention on two different cases:
\begin{itemize}
    \item The usual space $R^{n}$,  when $n$ is no more a positive integer, but it is a 
    real number \cite{Zinn}.
    \item The space of matrices $n \times n$, for $n$ real. In this case 
    we are mainly interested to study in the limit $n \to 0$ \cite{EA}.
\end{itemize}

The strategy followed by physicists in defining these strange objects consists in picking some 
properties of {\sl bona fine} spaces for positive integer $n$ and in making the analytic 
continuation to non integer $n$, without paying to much attention to the definition of the objects 
they are using.

A clarification of the mathematical structures involved would be extremely interesting because these 
spaces are extremely useful and their introduction has a strong heuristic value: it allowed us to 
obtain many results that later on have been proved using conventional techniques.  It is unlikely 
that this task can be done without generalizing the notion of space in an appropriate way.

\section{The case of $R^{n}$ with non-integer $n$}

One of the physical motivations for introducing spaces with non integer dimension is the 
following.  There are some physical systems whose properties cannot be computed directly 
in the case of the three dimensional space $R^{3}$, however  one can formally compute these properties 
in the space $R^{n}$ (that has dimension $n$) for $n$ near to some specific value (e.g. 
$n=4-\epsilon$ \cite{Zinn,Parisi}.  

The procedure used by physicists for introducing non-integer dimensional spaces is very simple.  The 
properties we would like to compute can be expressed in terms of some integrals.  In some cases  
the integrals can be computed for any dimension and their value can be extended to a meromorphic 
function of the dimensions \cite{ETI}.

Let us consider a very simple example: the volume of the $d$-dimensional unitary sphere is given by
\be
V(d)={\pi^{d/2} \over \Gamma(d/2+1)} \ . \label{SFERA}
\ee
The previous formula for $V(d)$  can be extended to an integer function of $d$ for real (and 
complex) $d$. Can one say that for generic real $d$ the volume of the unitary sphere is given by $V(d)$?

If an unitary sphere in a space with an non-integer dimension would exist, it is likely that its volume would 
be given by eq. (\ref{SFERA}). Unfortunately a space with an non-integer dimension is not defined 
and therefore the question of the volume of the unitary sphere is ill-posed. 

Now we could try to reverse the question.  We would like to say that a space with non-integer 
dimensions $d$ is characterized by the property that the volume of the unitary sphere is given by 
eq.  (\ref{SFERA}).  Obviously one integral is not sufficient to identify a space and other 
integrals must be defined (for example the volume of two or more intersecting spheres).  On the 
other hands this construction is likely to be empty; for example for $d=-3$, $V(d)=-2 \pi$ is 
negative and this does not make sense if the integration measure is positive \footnote{As we shall 
see later, negative dimensional spaces can be interpreted as positive dimensional spaces where some 
of the coordinates are anticommuting and the integration measure $d^{d}x$ is not positive definite 
\cite{PSC}.}.

The program may be the following; one can define geometrical objects (e.g. spheres.  lines, planes) 
only giving some rules on how to compute the measure of their intersections, without making any 
reference to the coordinates of the space where they are embedded.  If the construction is sufficiently 
rich, we could recover most of the geometrical results without having to introduce the space.  
Nothing forbids that some geometrical objects have a negative measure as far as they do no correspond 
anymore to real objects in a real space.

In the following I will firstly present a tentative  construction of the space with non integer 
dimensions along  these lines.

\subsection{The construction of a non-integer dimensions space}

The first objects I would like to define are the functions of one (or more) vectors that 
belong to a space with non-integer dimensions. This is standard construction in physics and 
it has been discussed with full mathematical rigour by Etingof \cite{ETI}.

In the usual integer dimensional case there is no difficulty 
in defining a mapping from $R^{n}$ to the real numbers. If $n$ is non-integer there is no 
reasonable way to define a generic point $x$ that belongs to this space, so  
we cannot define a  function $F(x)$ and the question of computing 
\be
\int d^{n}x \  F(x) \label{INTE}
\ee
does not make sense if $n$ is non-integer.

An alternative approach consists in giving some {\sl minimal} definition of the function $F(x)$ and 
defining $\int d^{n}x \  F(x)$ as a linear functional on the appropriate space without 
entering in the definition of $x$.

Let us consider a simple example of this strategy. We consider a function $F(x)$ of the form
\be
F(x)=f(x^{2}) \ . 
\ee
In this case for integer dimensions we have that 
\be
\int d^{n}x\  F(x)= n V(n) \int_{0}^{\infty} ds s^{n-1} f(s^{2}) \ , \label{INTZERO}
\ee
The previous expression defines a linear functional over the functions $f$ that is well defined if 
$f(s)$ goes to zero sufficiently fast when $s$ goes to infinity.  The extension to non integer $n$ 
presents no difficulty and one obtains a meromorphic function of $s$ if the function $f$ is 
$C^{\infty}$ at the origine.

The previous case may seems rather peculiar, however we can extend it to the case  where the
function $F$ depends on the vectors $x$ only via  scalar products with other vectors ($y$).
The whole construction should be  invariant under the rotation 
group $O(n)$, if we rotate the vectors $x$ and $y$.

The simplest case are   functions of two vectors and three vectors that 
respectively have the following form:
\be
G(x,y)=g(x^{2},y^{2}, x \cdot y) \ ,
\ee
and
\be
H(x,y,z)=h(x^{2},y^{2},z^{2}, x \cdot y, x \cdot z, y \cdot z) \ .
\ee
The integral over $x$ of the previous functions is obviously  well defined for integer $n$ and we will show 
that the final formulae are defined also for non-integer $n$.

Let us consider the following expression
\be
\int d^{n}y \ g(y^{2},x\cdot y)\ ,
\ee
evaluated for $x^{2}=r^{2}$. In the usual case, going to radial coordinates, one finds that
\be
\int d^{n}y\  g(y^{2},x\cdot y)= n V(n-1) \int_{0}^{\infty} ds s^{n-1}\int_{-\pi}^{\pi} d \theta 
\sin(\theta)^{n} f(s^{2},sr \cos(\theta)) \ . \label{INTUNO}
\ee
The previous expression makes sense also for 
non-integer $n$. In many cases it is a meromorphic function of $n$ and this allow us to define the 
r.h.s. of equation (\ref{INTUNO}) also in the region where the integral is not convergent \cite{ETI}.

This construction can be generalized to more complex cases; however an alternative and simpler 
construction, that gives the same results, is the following.  We suppose that the function $g$ has 
a generalized Laplace transform and that it can be written as:
\be
g(y^{2},x\cdot y)= \int_{0}^{\infty}  d\alpha \int_{0}^{\infty}  d\beta \ G(\alpha,\beta) 
\exp( -\alpha y^{2} -\beta x\cdot y) \ .
\ee
Using standard Gaussian integrals one finds that 
\be
\int d^{n}y \ g(y^{2},x\cdot y)=\int_{0}^{\infty}  d\alpha \int_{0}^{\infty}  d\beta G(\alpha,\beta) 
\left({\pi \over\alpha} \right)^{n/2}\exp\left( {\beta^{2}r^{2}\over 4 \alpha}\right) \ .
\ee
In the same way we may easily compute  integrals of the form 
\be
\int d^{n}y \ g(y^{2}, y\cdot x_{1}, \ldots ,y\cdot x_{k}) \ . \label{INTDUE}
\ee

Generally speaking a function of $k$ vectors  depends on the $k(k+1)/2$ invariants 
\footnote{For integer $n$ and sufficiently large $k$ the $k(k+1)/2$ invariants are not independent 
quantities, however this fact does not concern us directly, because we are considering a generic 
dimensional space.}.  In other words $f$ is a {\sl bona 
fide} function of one variable, $g$ of three variables and $h$ of six variables.  Only in 
our perverse imagination the arguments of the functions $F$, $G$ and $H$ are vectors in a 
non-integer dimensional space.
The crucial point where the dimension of the space enters is when we  attach a 
meaning each term of the expression {\sl integral of a function over the whole space}. 

We notice that the integral of a positive function is not {\sl a priori} positive, as can be seen 
from formula (\ref{INTUNO}).  Indeed the volume of the unit sphere is no more a positive function 
for negative dimensions.  For example negative values for the integral (\ref{INTDUE}) may arise if the $k$ 
vectors $x_{i}$ ($i=1,k$) form an orthonormal base and the value of $k$ is larger that the dimension 
of the space.  In other words we can always decompose an $n$ dimensional space in the product of an 
usual $k$ dimensional space (with integer $k$) with a space of dimension $n-k$.  If $n-k$ is 
negative the corresponding measure may be no more positive definite.  In this case, if 
we denote $y_{k}=x_{k}\cdot y$, we get:
\be
\int d^{n}y \ g(y^{2}, y_{1}, y_{2}, \ldots  y_{k})=(n-k)V(n-k)
\int dy_{1}dy_{2}\ldots dy_{k}dr r^{n-k-1}
g(r^{2}+\sum_{i=1,k}y_{i}^{2},y_{1},y_{2}\ldots y_{k})
\ , \label{INTTRE}
\ee
where the integral is done on the region  $r\ge 0$ and it should be understood in distribution 
sense as analytic continuation from the convergence region.  In the familiar case $k=n$ we recover the 
usual formula:
\be
\int d^{n}y \ g(y^{2}, y_{1}, y_{2}, \ldots  y_{k})=
\int dy_{1}dy_{2}\ldots dy_{k}
g(\sum_{i=1,k}y_{i}^{2},y_{1},y_{2}\ldots y_{k})
\ .\label{INTQUA}
\ee

In a similar way we can define derivatives, Laplacians and so on.
Generally speaking we can study vector-valued functions, under the constraint that they 
transform in a covariant way under the rotational group: e.g.
\be
x_{\mu}f(x^{2})\ .
\ee

In other words we study our space using the filter of rotational invariance: we consider only 
rotational invariant (or covariant) functions and we define a set of transformations (e.g. integrals 
and derivatives) that depend analytically on a parameter $n$;  for integer values of $n$ these 
transformations coincide with the usual ones for functions defined on a $n$-dimensional space.  

In general the 
geometrical meaning of this construction is unclear.
The only case where there is a simple geometrical interpretation is the case of negative integer
dimensions. For example in the  case of $n=-2$ we have that
\be
\int d^{-2}x F(x)=-2 \pi \int d\theta d\ba{\theta} F(\theta) \ ,
\ee
where $\theta$ is an anticommuting variable.
Indeed in the case where $F(x)=f(x^{2})$ both integrals give
\be
-2 \pi {df \over dx^{2}}{\biggr|}_{x^{2}=0} \ .
\ee
Negative even dimensional spaces can be interpreted as spaces with anticommuting (Grassmann) 
coordinates as it was noticed a long time ago \cite{PSC}.

\section{Zero-dimensional matrices}
\subsection{The replica method}
The interest in zero-dimensional matrices has been triggered by the so called replica method 
\cite{EA} for disordered systems.  

In many problems of statistical mechanics of random system we have to compute the quantity
\be
f=\lim_{N\to \infty} f(N) \ ,
\ee
where
\be
-f(N)=N^{-1}\int d\mu(J) \ln(Z_{J}(N))\ ,
\ee
Here the partition function $Z_{J}(N)$ is usually a positive quantity (defined in some 
complicated way) and $d\mu(J) $ is a normalized measure ($\int d\mu(J) $). 
For example in the Sherrington-Kirkpatrick model for spin glasses the partition function is given by
\be 
Z_{J}(N)=\sum_{ \{ \sigma\} }\exp\left( \beta 
\sum_{i,k=1,N}J_{i,k}\si_{i}\si_{k} \right)  \ ,
\ee
where the sum over $\{ \sigma\}$ denotes the sum over the $2^{N}$ configurations of the $N$ variables 
$\sigma_{i}$ that may tale only the values $\pm 1$.  The measure $d\mu(J) $ is a Gaussian measure 
over the $N(N+1)/2$ components of the symmetric $J$ matrix whose elements are identically 
distributed uncorrelated Gaussian  variables with zero average and variance $ N^{-1}$.

In the usual approach it is convenient to define the quantity
\be
-\Phi(N,n)= (Nn)^{-1}\ln\left(\int d\mu(J) (Z_{J}(N)^{n} \right).\label{Z}
\ee
As it can be trivially seen, if the appropriate integrals are convergent,
$\Phi(N,n)$ is an analytic function of $n$, that 
is regular around $n=0$, and 
\be
f(N)=\lim_{n \to 0}\Phi(N,n) \ .
\ee
This baroque construction has been introduced because in some problems for integer 
$n$ one can perform the integration over the variables $J$ and one can prove the following exact 
formula \cite{EA,SK}:
\be
\exp (-Nn\Phi(N,n))=\int dQ \exp(-NF_{n}(Q))\ , \label{EXACT}
\ee
where the integral is done over all $n \times n$ matrices $Q_{a,b}$ ($a,b=1,n$).
 such that  its diagonal elements are equal to 1 (i.e. $Q_{a,a}=1$).
Using the method of the point of maximum one can rigorously prove that for integer $n$
\be
\phi(n)\equiv\lim_{N\to\infty}\Phi(N,n) = F_{n}(Q^{*}) \ ,
\ee
where the matrix $Q^{*}$ minimizes the function $F_{n}(Q)$.

Forgetting physical motivations the mathematical problem is the following: given a function 
$F_{n}(Q)$ defined on $n \times n $ matrices one should go through the following steps:
\begin{enumerate}
    \item  To prove that the quantity $\Phi(N,n)$ defined in eq. (\ref{EXACT}) 
    can be extended to an analytic function of $n$.
    \item  To prove that the limit $f(N)=\lim_{n \to 0}\Phi(N,n)$ exists.
    \item  To prove that the quantity $f(N)$ has a limit ($f$) when $N$ goes to infinity.
    \item  To give a closed expression for the quantity $f$.
 \end{enumerate}
 
Of course everything depends on the form of the function $F_{n}(Q)$. For example in the case 
 where 
\be
    F_{n}(Q)=\frac\beta2 \Tr (Q^{2})
\ee
everything is trivial.
One explicitly finds that 
\be
\Phi(N,n)=-{n-1\over 2}\ln \left({\beta\over 2 \pi }\right)
\ee
and all the issues on the dependence on $n$ and $N$  of $\Phi(N,n)$ can be easily settled.

If the function $\Phi(N,n)$ has also the representation eq. (\ref{Z}), points 1 and 2 are easy, 
points 3 is more difficult, but it can be solved in some interesting cases \cite{GUERRA}. At the 
present time points 4 does not 
have a rigorous mathematical solution in the general case,

In order to gave an idea of the problems physicists are interested, it is convenient to write down 
some of the functions that have been studied in the literature. 
\bea
F_{n}(Q)= -\Tr [\ln(Q)] + \beta^{2}\Tr (Q^{2}) \\
F_{n}(Q)=-\Tr [\ln(Q)]  +\beta^{2}\sum_{a,b}(Q_{a,b})^{3}\\
F_{n}(Q)=-G_{n}(Q) +\beta^{2} \Tr (Q^{2})  \nonumber
\eea
where in the last case (that correspond to the SK model, where we have that:
\be
\exp (n G_{n}(Q)) =\prod_{a=1,n}\left[\sum_{\sigma_{a}=\pm 1}\right] 
\exp\left( \sum_{a,b}Q_{a,b}\sigma_{a}\sigma_{b}\right) \label {G}
\ee 

The  physicists have used the following approach.  For integer $n$ the  matrix $Q^{*}$ 
that minimizes the function $F_{n}(Q)$ can be easily found and in this way one computes $\phi(n)$ 
for integer $n$.  Usually $\phi(n)$ has a simple expression that can be 
analytically continued to $n=0$.  We would naturally guess that this analytic continuation gives the 
exact result.  

However in general this is not the correct procedure.  The function $\phi(n)$ is defined as the 
infinite $N$ limit of $\Phi(N,n)$ and, although $\Phi(N,n)$ is analytic in $n$, the function $\phi(n)$ 
may be a non-analytic function of $n$ (e.g. it may have a singularity in the interval 0-1) so that 
the analytic continuation of $\phi(n)$ is not the limit for $N$ going to infinity of $\Phi(N,n)$.  
In some crucial cases is was proved that this happens (the naive approach gives inconsistencies) and 
something else should be tried \cite{mpv},

The next approach consists in trying to use the method of the point of maximum directly for non 
integer $n$, with all possible complications that may arise in dealing directly with functions 
defined on a non integer dimensions case \cite{mpv}.  This approach is consistent (in the sense that the final 
results are not contradictory).  Its results have not yet been proved to be correct and goes under 
the name of spontaneously broken replica symmetry (the naive approach is usually called the replica 
symmetric one).

It is a great challenge to people working in probability theory to prove (or to disprove) that the 
results of the replica broken approach are correct \cite{TALAGRAND}. 
 
\subsection{The replica symmetry}

We have already seem that symmetry considerations play a crucial role in defining quantities on 
non-integer dimensional spaces.  It is therefore reasonable to investigate the symmetry properties 
of our problems.  Moreover the presence of a symmetry group in a function $F$ allow us to divide the 
set of its stationary points into orbits under the action of symmetry group.

Let us introduce the usual permutation group $S_{n}$ (i.e. the replica group) and let us denote by 
$\pi(a)$ (for $a=1,n$) the element where $a$ is carried by the permutation $\pi$.  The permutation 
group (that is a subgroup of the rotational group) acts in a canonical way on the matrices $Q$, 
permuting both row and columns:
\be
Q^{\pi}_{a,b}=Q_{\pi(a),\pi(b)} \ . 
\ee

In the cases we are interested the permutation group is a symmetry of the 
problem, i.e. $F(Q^{\pi})=F(Q)$. This symmetry  is related to the procedure  used in the 
construction of  the function $F$  it is clearly exact in the examples we have considered.

If a function $F$ has a symmetry, the symmetric point is a natural candidate for a minimum: it 
is automatically a stationary point, i.e. a solution of the equation
\be
{\partial F \over \partial Q_{a,b}} =0 \ . \label{STA}
\ee
A  matrix that is left invariant by the 
action of the permutation group must have all off-diagonal elements equal:
\be
Q_{a,b}=q \ \ \ \ \mbox{for} \ \ \ a\ne b \ . 
\ee
(The diagonal elements of $Q$ are equal to one by construction.)

However the symmetric solution of equation (\ref{STA}) does not need to be a minimum, it may be a 
maximum, or a saddle.  In order to establish the nature of the symmetric solution we have to compute 
the eigenvalues of the Hessian
\be
{\cal H}_{a,b;c,d}={\partial^{2 }F \over \partial Q_{a,b} Q_{c,d}} \ .
\ee
In one of the most interesting cases (the Sherrington-Kirkpatrick model for spin glasses \cite {SK}, 
i.e. the last of our  examples) one finds that the eigenvalues of the Hessian are all positive 
for positive integer $n$, their analytic continuation is still positive for $n>n_{c}$, but some of 
them change sign at $n_{c}$ where $0<n_{c}<1$ \cite{DAT}.  By changing the parameter $n$ a minimum 
becomes a saddle point and an other stationary point becomes the true minimum: this a very usual phenomenon (e.g. a 
bifurcation or spontaneous symmetry breaking) in an unusual setting.

\subsection{Breaking the replica symmetry} 

If the symmetric point is not a minimum, we can look for a minimum that is not 
symmetric. In this case physicists say that the symmetry group is spontaneously broken. The 
subgroup that leaves invariant the true minimum (the little group) is the surviving (or unbroken) symmetry group.

Our task is to find a new candidate for a minimum.  Different attempts have been done.  
The most promising candidate  \cite{mpv,ParisiW} is given by the 
matrix $Q$, whose off-diagonal elements can be written as
\bea
Q_{a,b}=q_{1} \ \ \ \ \mbox{if} \ \ \ I(a/m)=I(b/m) \ , \\
Q_{a,b}=q_{0} \ \ \ \ \mbox{if} \ \ \ I(a/m)\ne I(b/m)  \label{ONESTEP} \ ,
\eea
where $I(x)$ is the integer part of $x$ and $m$ is an integer that divides $n$.  Obviously if 
$q_{1}=q_{0}=q$ we recover the previous case.  The symmetry group that leaves invariant this matrix 
is (for $q_{1}\ne q_{0}$) the semidirect product of $S_{n/m}$ with the direct product of $n/m$ 
copies of $S_{m}$ and it is obviously a subgroup of $S_{n}$.

The value of $F(Q)$ evaluated at this saddle point  gives a function $F(q_{0},q_{1},m)$, 
that can be written in a relatively simple form, depending on the model
(in the worst case it involves a few integrals).

Now a natural question is which values of $m$ are  allowed. Originally $m$ should be 
an integer that divides $n$, however, when $n$ is no more an integer, there is apparently 
no reason for imposing that  $m$ remains an integer.  It is clear that we are in a 
nobody land and any prescription is  arbitrary. We will present 
the {\sl standard}  prescription, that may justified {\sl a posteriori} by its success in giving 
sensible results, that are quite likely the correct ones. We will follow the rules:
\begin{itemize}
    \item Non-integer values of $m$ are allowed as soon as $n$ is no more a positive 
    integer.
    \item The interesting region for $m$ is $1\le m \le n$ for $n>1$ and $n\le m \le 1$ 
    for $n<1$.
    \item We look for a stationary point that satisfies the condition that all the eigenvalues of ${\cal 
    H}$ are positive in order to 
    find the minimum of $F(Q)$.  In the previous defined interesting region ($n\le m \le 1$ 
    for $n<1$) this condition implies that $F(q_{0},q_{1},m)$ is a maximum as function of 
    $q_{0}$ and $q_{1}$.   This confusing situation is due to the fact the if the number of dimensions 
    is negative, the two conditions (${\cal H }$  has non negative eigenvalues and
    $(x, {\cal H } x)\ge0$) are not equivalent.
    \item By uniformity we decide that the function $F(q_{0},q_{1},m)$ 
    should also maximized with respect to $m$.  The final prescription consists in finding the 
    maximum with respect all the three arguments.
\end{itemize}

This prescription is called one step replica symmetry breaking.  In certain models it is 
believed to give the exact results. i.e. in the cases for which one obtains a non-negative 
Hessian (an alternative approach to these models is sketched in the appendix).  In other 
cases (e.g. in the SK model) one does not obtain a non negative value of the Hessian 
although the absolute value of the negative eigenvalue decrease be a large factor (around 
10). The prescription we have introduced is apparently able to compute in an exact way 
the quantity $f$ in some models. In the other models, where it does not give a consistent result,
it is a step toward the correct solution, that will be found in the next section.

\subsection{Full replica symmetry breaking}

As we have mentioned before the one step replica symmetry scheme does not produce consistent 
results in the case of the SK model.  If we pay attention to the scheme that we have used 
to break the permutation group, the statements of the previous section corresponds to 
saying that in the limit $n \to 0$ the unbroken subgroup of $S_{n}$ is the semidirect product of the 
group $S_{0}$ with the direct product of zero copies of the group $S_{m}$.  In other words 
the group $S_{0}$ contains itself as a subgroup and it is therefore an infinite group (it 
also contains infinitesimal elements \cite{PASLA}).  This fact may be not surprising if we 
ponder the fact the group $S_{0}$ is obtained as analytic continuation of all the $S_{n}$ 
groups, with arbitrary $n$.

One can  generalize  the formula eq.  (\ref{ONESTEP}) to the case where there 
are $k+1$ parameter $q_{k}$ and $k$ parameter $m_{k}$. The previous formulae correspond to the case 
$k=1$.
We present the explicit construction in the  case $k=2$. We assume the $n$ is a multiple of $m_{2}$ 
and $m_{2}$ is a multiple of $m_{1}$.
The off-diagonal elements of the matrix $Q$,   can be written as
\bea
Q_{a,b}=q_{2} \ \ \ \ \mbox{if} \ \ \ I(a/m_{2})=I(b/m_{2}) \ , \\
Q_{a,b}=q_{1} \ \ \ \ \mbox{if} \ \ \ I(a/m_{2})\ne I(b/m_{2})  \ \ \ \ \mbox{and} \ \ \ 
I(a/m_{1})= I(b/m_{1})
\label{TWOSTEP} \ , \\
Q_{a,b}=q_{0} \ \ \ \ \mbox{if} \ \ \ I(a/m_{1})\ne I(b/m_{1})  \ ,\nonumber
\eea
It is remarkable that after a permutation one could write  
write such a matrix under the form $Q_{a,b}=Q(a-b)$, where the function $Q(a)$ is a simple 
function of $a$ \cite{PS}. Indeed this happens if  we set
\bea
Q(a)=q_{2} \ \ \ \ \mbox{if} \ \ \ n/m_{2} \ \ \ {\rm divides} \ \ a  \ , \\
Q(a)=q_{1} \ \ \ \ \mbox{if} \ \ \ n/m_{1} \ \ \ {\rm divides} \ \ a  \ \ \  {\rm  and } \  \ 
n/m_{2} \ \ \ 
{\rm does \ not \ divide} \ \ a, \\ 
Q(a)=q_{0} \ \ \ \ \mbox{if} \ \ \ n/m_{1} \ \ {\rm does \ not \ divide} \ \ a, \nonumber
\eea
It is not clear if this translational invariant formulations, that is at the heart of 
the $p$-adic approach \cite{PS}, has  a deeper meaning (up to this moment it has been 
used only at the technical level).

Also in this case we first compute the function $F$ for integer $n$, $m_{1}$ and 
$m_{2}$ and we analytically continue the result up to $n$=0 \footnote{The group properties of 
this solution are discussed in details in \cite{HUN}.}.  At the end of the computation
we do not require any more 
that the $m$'s are integer.   The final expression is maximized as function of the $q$'s and 
of the $m$'s, that turn out to be in the region $0<m_{1}<m_{2}<1$. In the case where the one step solution is not 
good, (i.e. negative Hessian) also the two steps solution is not good (the Hessian is 
still negative).

It is clear that this construction can go on for any number of steps we want in a recursive 
way.  A detailed computation shows that we always find the the Hessian is negative (for 
example in the SK model near the critical temperature the minimum eigenvalue of the 
Hessian is $-A(2k+1)^{-2}$ with positive $A$).  It seems reasonable that the interesting case correspond to 
the limit $k \to \infty$.

In order to control better the $k \to \infty$ it is important that for finite $k$ and in the limit $n \to 0$ the 
interesting solutions
of the stationary equation 
$
\partial F / \partial Q_{a,b} 
$
are always in the region 
\be
0\equiv m_{0}<m_{1}\ldots <m_{k}<1\equiv m_{k+1} \ .
\ee
One can thus associate to the matrix $Q$ a piecewise constant function $q(x)$ such that
\be
q(x) =q_{i} \ \ \mbox{for}  \ \ \ \ m_{i}<x<m_{i+1} \ .
\ee
If we send $k$ to infinity, one finds by an esplicite computation that the function $q(x)$ becomes a 
continuous function. In this way we have a canonical way to associate to a function $q(x)$ a 
matrix in the {\sl space} of zero by zero matrices.

The introduction of the function $q(x)$ is not so arbitrary, Indeed simple expressions can be 
obtained for the $S_{n}$-invariants in the limit of $n$ going to zero also for finite $k$.  I present 
just two simple  examples:
\bea
\lim_{n \to 0} n^{-1}\Tr (Q^{2})= 1-\int_{0}^{1}dx q(x)^{2},\\
\lim_{n \to 0} n^{-1}\Tr (Q^{3})= 1- 3\int_{0}^{1}dx q(x)^{2}  +\frac32 \int_{0}^{1}dx\int _{x}^{1}dy q(x)q(y)^{2}
+\frac12
\int_{0}^{1}dx x q(x)^{3}\ .
\eea
The last formula is related to the formula for the volume of the intersection of 
three $p$-adic spheres \cite{PS}.

A more complex example is given by the quantity $G(0)$, defined in eq. (\ref{G}): a non 
trivial computation 
implies that
\be
G(0)=f(0,0) \ ,
\ee
where the function $f(x,h)$ is defined in the    strip $0\le x\le 1$, $-\infty < h 
<\infty$ and it satisfies the following  non-linear differential equation :
\be
{\partial f \over \partial x}=-\frac12 {dq\over dx}
\left( {\partial^{2}f \over \partial h^{2}} +x \left({\partial f \over \partial 
h}\right) ^{2}\right) \ .
\ee
the with the
boundary conditions 
\be
f(1,h)=\ln(\cosh(\beta h)).
\ee

In all the previous cases the end of the computation the free energy becomes a functional of the 
function $q(x)$ ($F[q]$) and this functional has an explicit expression that in some cases 
may be rather complex.  Maximizing the functional $F[q]$ (and depending on the temperature 
range this can be done or analytically or numerically) one finds a consistent solution of 
many models (including the SK model) such that all the infinite eigenvalues of the Hessian 
\footnote{The Hessian becomes now a rather complex integral operator and it has both an 
infinite point spectrum and a continuous spectrum.} are non negative.  Moreover this 
solutions satisfies all inequalities that one derive from general principles (e.g. 
positivity of the entropy).

It obvious that all this does not make sense from a strict mathematical point of view because the 
prescriptions we have used  are somewhat arbitrary, although they  look reasonable. On 
the other ends the replica method has been very successfully as can be seen in the next 
section.

\section{The success of the replica method}

In the case of the Sherrington Kirkpatrick model the first success of the replica method, especially 
in the case of full replica symmetry breaking, is the existence of a consistent solution (a result that 
is 
not evident {\sl a priori}).

It is also remarkable that in all the many infinite range models, where the replica method 
has applied, the results are in a very good  agreement with the computer simulations and 
they coincide with the exact results, when available.

It is very interesting that one can rederive most of the results of the replica method, without 
using replicas, but doing probabilistic arguments.  It appears that the form of the matrix $Q$ codes 
some information on the probability distribution of the equilibrium configurations of the system in 
a way that it would take too much time to explain \cite{mpv,ParisiW}.  We only say that the 
organization of the pure states of the systems (called lumps by Talagrand \cite{TALAGRAND} in this 
context) mirrors the form of the matrix $Q$: e.g. the ultrametric form of the matrix $q$, that is a 
remnant of its $p$-adic origine, implies that the distances among states satisfy the ultrametric 
inequality.  This alternative probabilist approach is not evident and it was found a few years after 
the establishment of the replica approach.  Moreover in some models non-trivial rigorously results 
have been obtained that point toward the correctness of the results obtained via the replica method 
\cite{TALAGRAND}.

Most recently in a very clever paper Guerra have been able to prove that in the SK model (and in 
similar models) the following inequality si satisfied:
\be
f \ge F[q] \ \ \ \forall q(x) \ .
\ee
This result is important because it shows that the functional $F[q]$, that is obtained via the 
replica method, is related to the true value ($f$) of the free energy.

The same results of the replica method may be obtained in the framework of off-equilibrium 
dynamics that has been shown to be equivalent to the replica method \cite{CK,FM}.  Recent 
experimental results \cite{EX} beautifully confirm the theoretical predictions of this approach.

The existence of alternative routes clearly shows that the results make sense.  
In any case  the replica method has a very strong heuristic value and it 
would be surprising if we could not assign a precise mathematical meaning to such an 
useful method.

\section*{Appendix}
In this appendix we report an explicit computation in the case of one step replica
symmetry breaking that may have some chance to be the starting point of a rigorous
mathematical approach \cite{CPV}. As we shall see, this result suggests that there is a 
correspondence from a class of $ 0 \times 0$ matrices with an infinite set of $n \times 
n$ matrices for any $n$.

At the technical level the idea is relatively simple. We write for negative $s$
\be
\Gamma(-s){\ba{Z^{s}} }=\int_{0}^{\infty} dt t^{-1-s} {\ba{\exp (-t Z)} }=
\int_{0}^{\infty} dt t^{-1-s} \sum_{n=0,\infty} {(-t)^{n}\over n!} {\ba{Z^{n}} }  \  , \label{DERRIDA}
\ee
where everything has an implicit dependence on $N$ and the bar denotes the average over $J$.  If we 
continue the previous formula up to $s \to 0$ we obtain and expression for $\ba{\ln(Z^{s})} $

If we can control the expression for
$\ba{Z^{n}} $ with sufficient accuracy when $N \to \infty$, we have done our job.  This
task is not easy, because of possible cancellations due to the minus signs: the
leading term of the sum may be not the sum of the leading terms for $\ba{Z^{n}} $ and 
therefore we have to control $\ba{Z^{n}} $ with a very high accuracy.

We have already seen  that the quantity 
$\overline{Z^{n}}$ has an exact expression that is given by eq.  (\ref{EXACT}).  An 
interesting conjecture is that we obtain the asymptotically correct result for $\overline{ 
Z^{s} }$, if inside eq. (\ref{DERRIDA}) we make the approximation
\be
{\ba{Z^{n}} }\approx \sum_{Q^{*}} \exp (-N F_{n}(Q^{*})) \ ,
\ee
where the sum is done over all the $n \times n$ matrices $Q^{*}$ that are local minima
of $F_{n}(Q)$. This approximation  obviously gives the asymptotically dominant result for 
$\ba{Z^{n}} $, where only the global minimum counts, however it is not evident that it gives 
the correct result when it is used for computing $\ba{Z^{s}} $.
   
In the general case we do not know if conjecture is true, how to compute
all the mimima and if this
conjecture does reproduce the result of the replica approach.  However in some models
where replica symmetry  is broken at one step, the appropriate computations can be
done \cite{CPV} and one finds the same result obtained with the replica method.

There is indeed a class of models where for positive integer $n$ all the minima of $F[Q]$ 
can be obtained in the following way: one divides the indices in $M$ sets of size $m_{k}$ 
($k=1,M$) where $\sum_{k}m_{k}=n$. One sets
\be
Q_{a,b}=q_{k} \ ,
\ee
if both $a$ and $b$ belongs to the $k^{th}$ set; otherwise one has $Q_{a,b}=0$.  In some cases 
\cite{CPV} one finds that if  the leading 
contribution (that coming from the global minimum) for each $n$ is used, the result coincides with 
the one coming form the replica symmetric approximation (that for a certain range of the 
parameters is not correct).  One the other end, if one consider the contribution coming from all 
possible minima (that are subdominant for give $n$), at the end one recovers the replica broken 
result where one has to find the maximum of the free energy for $m$ in the region $0\le m \le 1$.

We have no idea of how the afore-mentioned conjecture can be proved and it would be 
interesting to test it in a wider context. We have mentioned it here because it  
hints that the  matrices for non integer $n$ and $m$ used in the replica method  are
related to families of matrices that are defined for all integer $n$.

\end{document}